\documentclass[12pt, a4paper]{article}
\usepackage[cp1251, ctt]{inputenc}
\usepackage[T2A]{fontenc}
\usepackage{amsthm,amsfonts,amssymb,cite}
\usepackage[english]{babel}
\usepackage[tbtags]{amsmath}
\usepackage{graphicx}

\def\Tr{\operatorname{Tr}}

\let\cal\mathcal

\DeclareSymbolFont{bletters}{OML}{cmm}{bx}{it}
\DeclareMathSymbol{\bla}{\mathord}{bletters}{'025}
\DeclareMathSymbol{\bmu}{\mathord}{bletters}{'026}
\DeclareMathSymbol{\bth}{\mathord}{bletters}{'022}
\DeclareMathSymbol{\bfI}{\mathord}{bletters}{"49}
\DeclareMathSymbol{\bdl}{\mathord}{bletters}{"0E}
\DeclareMathSymbol{\bDl}{\mathord}{bletters}{"001}

\def \si{\sigma}
\def \la{\lambda}
\def \be{\beta}
\def \ga{\gamma}
\def \ta{\theta}
\def \dl{\delta}

\def \Dl{\Delta}
\def \Ga{\Gamma}
\def \CA{\mathcal A}
\def \CM{\mathcal M}

\def \CV{\mathcal V}

\def \CC{\mathcal C}

\def \CP{\mathcal P}

\def \CT{\mathcal T}

\def \CI{\mathcal I}
\def \BC{\mathbb{C}}
\def \BR{\mathbb{R}}
\def \BZ{\mathbb{Z}}
\def \BI{\mathbb{I}}

\begin{document}

\title{
$${}$$
{\bf
The Ising limit of the XXZ~Heisenberg magnet and certain thermal correlation functions}}
\author{
$${}$$
{\bf\Large N. M. Bogoliubov$^\dagger$, C. Malyshev$^\ddagger$}\\[0.5cm]
{\it\small Steklov Mathematical Institute, St.-Petersburg Department,
RAS}\\
{\it\small  Fontanka 27, St.-Petersburg, 191023, Russia} \\
[0.5cm]
$^\dagger$ e-mail: {\it bogoliub@pdmi.ras.ru}\\
$^\ddagger$ e-mail: {\it malyshev@pdmi.ras.ru}
}

\date{}

\maketitle

\vskip1.0cm

\rightline{\emph{{\it In Memorium of A. N. Vassiliev}}}

\vskip1.0cm

\renewcommand{\abstractname}{\centerline{\normalsize
{\bf{Abstract}}}}
\begin{abstract}
\noindent The  spin-{\it 1/2} $XXZ$ Heisenberg magnet is considered for the case of the anisotropy parameter tending to infinity (so-called, {\it Ising limit}). A thermal correlation function of the ferromagnetic string is calculated over the ground state. The approach to the calculation of the correlation functions is based on the observation that the wave function in the limit considered is expressed in terms of the symmetric Schur functions. It is demonstrated that at low temperatures the amplitude of the asymp\-to\-tic\-al expression of the present  correlation function is proportional to the squared numbers of strict boxed plane partitions.
\end{abstract}

\leftline{\emph{{\bf Keywords:}}  Heisenberg magnet, Ising limit,
correlation functions}

\thispagestyle{empty}
\newpage

\section{Introduction}\label{tmf:sec1}

The Ising model was one of the favorite models of A. N. Vassiliev. His studies were concentrated on the scaling properties of this model \cite{vas1, vas2}. It is remarkable that in a special limit of the anisotropy parameter tending to infinity, the Hamiltonian of the spin{\it -1/2} Heisenberg $XXZ$ chain is transformed into the Hamiltonian of one-dimensional Ising model \cite{yy3}. This limit is usually called as the {\it Ising limit} of the $XXZ$ chain \cite{god}. In our paper
we shall investigate the Ising limit of the $XXZ$ magnet using an effective Hamiltonian which is commutative with the Ising Hamiltonian and has a complete system of the eigen-functions which is common with that of the Ising Hamiltonian. This effective Hamiltonian is an appropriate object for studying of the Ising limit since it is related to the transfer-matrix of the four-vertex model, which arises in the limit of infinite anisotropy from the transfer-matrix of the six-vertex model. In their turn, the trace identities relate the $XXZ$ Hamiltonian and the transfer-matrix of the six-vertex model.

The problem of calculation of the correlation functions of  $XXZ$ Heisenberg model in the framework of the Quantum Inverse Scattering Method \cite{f1} has required serious efforts \cite{ KBI1, KBI2, vk, vk1, ml1, ml2}.
We shall show that calculation of the correlation functions in the Ising limit is closely related to the theory of the symmetric functions \cite{macd}, and thus is closely connected with the problems of enumerative combinatorics, such as the
random walks of vicious walkers \cite{1, 3, 5, 8, 9} and enumeration of boxed plane partitions \cite{bres, b8}. In this paper we shall calculate the thermal correlation functions of the ferromagnetic string over the ground state using the approach developed in  \cite{b1, b3, b7, b4, b5}.

The $XXZ$ spin-{\it 1/2} Heisenberg magnet is defined on the one-dimensional lattice consisting of $M+1$ sites labeled by elements of the set $\CM\equiv\{0 \le k \le M,\,k \in\BZ\}$, \break $M+1=0\pmod{2}$. The corresponding spin Hamiltonian is defined as follows:
\begin{equation}
{\widehat H}_{\rm XXZ}=-\frac 12\sum_{k=0}^M (\si_{k+1}^{-}\si_k^{+} + \si_{k+1}^{+}\si_k^{-}+\frac\Dl 2\,(\si_{k+1}^z\si_k^z-1)+h\,\si_k^z)\,, \label{xxzham}
\end{equation}
where the parameter $\Dl\in\BR$ is the anisotropy of the model, and $h$ is the  external magnetic field. The local spin operators $\si^\pm_k = \frac12(\si^x_k\pm i\si^y_k)$ and $\si^z_k$, dependent on the lattice argument $k\in\CM$, are defined standardly (see \cite{KBI1, KBI2}),
and their commutation rules are given by the relations:
$
[\,\si^+_k,\si^-_l\,]\,=\,\dl_{k,l}\,\si^z_l$, $[\,\si^z_k,\si^\pm_l\,]\,=\,\pm 2\,\dl_{k,l}\,\si^{\pm}_l
$.
The state-space is spanned over of the state-vectors $\bigotimes_{k=0}^M \mid\!\! s\rangle_{k}$ , where $s$ implies either $\uparrow$ or $\downarrow$. The spin ``up'' and ``down'' states
($\mid\uparrow\rangle$ and $\mid\downarrow \rangle$, respectively) provide a natural basis in the linear space ${\BC}^2$ so that
\[
\mid\uparrow\rangle\equiv \Bigl(
\begin{array}{c}
1 \\
0
\end{array}
\Bigr)\,,\qquad\mid\downarrow\rangle\equiv \Bigl(
\begin{array}{c}
0 \\
1
\end{array}
\Bigr)\,.
\]
The periodic boundary conditions $\si^{\#}_{k+(M+1)}=\si^{\#}_k$ are imposed. The Hamiltonian (\ref{xxzham}) commutes with the operator of third component of the total spin ${\widehat S}^z$:
\begin{equation}
\lbrack{\widehat H}_{\rm XXZ}, {\widehat S}^z]\,=\,0\,,\qquad
 {\widehat S}^z\equiv \frac 12\sum_{k=0}^M\sigma_k^z\,.  \label{spi}
\end{equation}

To represent $N$-particle state-vectors, $\mid\!\Psi_N(u_1,\dots,u_N)\rangle$, let the sites with spin ``down'' states be labeled by the coordinates $\mu_i$ forming a strict partition $\boldsymbol{\mu}\equiv (\mu_1, \mu_2,\,\dots\,, \mu_N)$, where $M\geq\mu_1>\mu_2>\,\dots\,> \mu_N \geq 0$. There is a correspondence between each partition and an appropriate sequence of zeros and unities of the form: $\bigl\{e_k\equiv e_k(\boldsymbol{\mu})\bigr\}_{k\in \CM}$, where $e_k \equiv \dl_{k, \mu_n}$, $1 \leq n\leq N$.
The Hamiltonian (\ref{xxzham}) is diagonalised \textit{via} the following ansatz:
\begin{equation}
\mid\!\Psi_N({\textbf u})\rangle \,=\,\!\!
\sum_{\{e_k(\bmu)\}_{k\in\CM}}
\chi_{\bmu}^{\rm XXZ} ({\textbf u}) \prod\limits_{k=0}^M
(\si_k^{-})^{e_k} \mid \Uparrow \rangle\,,
\label{bwf}
\end{equation}
where the sum is taken over $C^N_M$ strict partitions ${\bmu}$, and $\mid\Uparrow\rangle \equiv \bigotimes_{n=0}^M \mid \uparrow \rangle_n$ is the fully polarized state with all spins ``up''. It is proposed to use bold-faced letters as short-hand notations for appropriate $N$-tuples of numbers: for instance, ${\textbf u}$ instead of $(u_1, \dots, u_N)$, etc. Therefore, the wave function $\chi_\bmu^{\rm XXZ} ({\textbf u})$ in (\ref{bwf}) is of the form:
\begin{eqnarray}
\chi_\bmu^{\rm XXZ} ({\textbf u})\,=\,\sum_{S_{p_1, p_2, \dots , p_N}} \CA_S({\textbf u})\,u_{p_1}^{2 \mu_1} u_{p_2}^{2 \mu_2} \ldots u_{p_N}^{2 \mu_N}\,,\label{xxzwf}\\[0.0cm]
\displaystyle{\CA_S ({\textbf u})\,\equiv\,\prod_{1\le j<i\le N}\frac{1-2\Dl u_{p_i}^2 + u_{p_i}^2 u_{p_j}^2}{u_{p_i}^2 - u_{p_j}^2}}\,,
\label{ampxxz}
\end{eqnarray}
where summation goes over all elements of the symmetric group
\[
S_{p_1, p_2, \dots , p_N}\equiv {\small S   \Bigl(\begin{matrix}1, & 2, & \dots , & N \\ p_1, & p_2, & \dots, & p_N \end{matrix} \Bigr)}\,.
\]
The state-vectors (\ref{bwf}) are the eigen-states of the Hamiltonian (\ref {xxzham}),
if and only if the variables $u_l$ satisfy the \textit{Bethe equations}
\begin{equation}
u_l^{2(M+1)}=(-1)^{N-1}\prod_{k=1}^N\frac{1-2\Dl u_l^2+u_l^2 u_k^2}{
1-2\Dl u_k^2 + u_l^2 u_k^2}\,, \qquad1\le l\le N.
\label{xxzbethe}
\end{equation}
The corresponding eigen-energies are given by
\begin{equation}
E_N({\textbf u})\,=\,- \frac12\sum_{i=1}^N(u_i^2+u_i^{-2}-2\Dl-2h)\,.
\label{een}
\end{equation}

The limit of zero anisotropy, $\Dl\to 0$, transforms (\ref{xxzham}) to the Hamiltonian of the $XX$ model on the periodic chain
\begin{equation}
\widehat H_{\rm XX}\,\equiv\,-
\frac{1}{2}\sum_{k=0}^M\,(\widehat h_{k+1,k}+h\,\si_k^z)\,,\qquad  \widehat h_{k+1,k} \equiv \si_{k+1}^{-}\si_k^{+} +
\si_{k+1}^{+}\si_k^{-}\,.
\label{anis1}
\end{equation}
This limit is known as the free fermion limit of the $XXZ$ chain, and it is the most studied one \cite{lieb, niem, iz11, mal1}.

In the next section II we shall consider the Ising limit corresponding to the limit $\Dl\to-\infty$.
The thermal correlation functions of the ferromagnetic string operator are obtained in Section III.
Section IV presents several results for the  four-vertex model which is the $\Delta\rightarrow -\infty$ limit of the six-vertex model. In Section IV it is demonstrated that the effective Hamiltonian commutes with the transfer-matrix of the
four-vertex model, the orthogonality of the eigen-functions is proved, and the relation between the four-vertex model and the boxed plane partitions \cite{macd} is discussed.
The low temperature asymptotics of the correlation function is studied in the Section V.
Discussion in Section VI concludes the paper.

\section{The Ising limit of the ${\rm\bf XXZ}$ chain}\label{tmf:sec2}

Less studied limit of the $XXZ$ model is the \textit{Strong Anisotropy} (SA) limit $\Dl\to-\infty$ \cite{god, alc, abar, ess1}. This limit is called the \textit {Ising limit} since the simplest Hamiltonian in this limit turns out to be the Hamiltonian of the one-dimensional Ising model \cite{god}:
\begin{equation}
\lim_{\Delta\rightarrow \,-\infty}\frac{1}{\Delta}\,{\widehat H}_{\rm XXZ}\,=\,\widehat H_{\rm Is}\,\equiv\,\frac{-1}{4}\sum_{k=0}^M(\si_{k+1}^z \si_k^z -1)\,.
\label{ising}
\end{equation}
However we shall study the strong anisotropy limit using the effective Hamiltonian $\widehat H_{\rm SA}$, which is formally equivalent to the $XX$ Hamiltonian (\ref{anis1}), provided the latter is supplied with the requirement forbidding two spin ``down'' states to occupy any pair of nearest-neighboring sites:
\begin{equation}
\widehat H_{\rm SA}\,=\,-
\frac{1}{2}\sum_{k=0}^M\CP\,(\widehat h_{k+1,k}+h\,\si_k^z)\,\CP\,, \qquad
\CP \equiv \displaystyle{ \prod_{k=0}^M}(\BI-\hat q_{k+1} \hat q_k)\,.
\label{anis2}
\end{equation}
The projectors $\CP$ ``cut out'' the spin ``down'' states for any pair of nearest-neighboring sites. The local projectors onto the spin ``up'' and  ``down'' states are of the form:
\begin{equation}
\hat q_k\,\equiv\,\frac12\,(\BI - \si^z_k)\,,\quad \check q_k\,\equiv\,\frac12\,(\BI + \si^z_k)\,,\quad \check q_k + \hat q_k\,=\,\BI , \qquad k\in\CM\,,
\label{proj}
\end{equation}
where $\BI$ is the unit operator in $(M+1)$-fold tensor product of the linear spaces~$\BC$. The point is that the Hamiltonians (\ref{ising}) and (\ref{anis2}) are commutative. Indeed, the Hamiltonian (\ref{ising}) can be represented by means of (\ref{proj}) as follows:
\begin{equation}
\widehat H_{\rm Is}\,=\,\widehat N\,-\,\sum_{k=0}^M\hat q_{k+1} \hat q_k\,,\qquad \widehat N\,\equiv\,\sum_{k=0}^M \hat q_k\,,
\end{equation}
where $\widehat N$ is the operator of number of particles commuting with the Hamiltonians $\widehat H_{\rm XX}$ and $\widehat H_{\rm SA}$. On another hand, the definition of the projector $\CP$ yields:
\[
\CP\sum_{k=0}^M\hat q_{k+1} \hat q_k=\sum_{k=0}^M\hat q_{k+1} \hat q_k\CP=0.
\]
Then, $[\widehat H_{\rm SA}, \widehat H_{\rm Is}]=0$ and so the Hamiltonians (\ref{ising}) and (\ref{anis2}) possess a common system of the eigen-functions.

The wave function (\ref{xxzwf}) takes the form in the limit $\Dl\to-\infty$:
\begin{equation}
\chi_\bmu^{\rm SA}({\textbf u})=\det(u_j^{2(\mu_k-N+k)})_{1\leq j, k\leq N}\,\prod_{1\leq n<l\leq N}(u_l^2-u_n^2)^{-1}\,,
 \label{xxwf1}
\end{equation}
where the coordinates of the spin ``down'' states form a strict decreasing partition $\bmu$, as in (\ref{bwf}).
It follows from (\ref{xxwf1}) that the wave function is not equal to zero if and only if the elements $\mu_i$, $1\le i\le N$, satisfy the \textit{exclusion condition}: $\mu_i>\mu_{i+1}+1$. In other words, in the considered limit occupation of nearest sites is forbidden, and the \textit{hard-core diameter} equal to duplicated inter-site separation arises.
The exponential parametrization  $u^2_j=e^{i\ta_j}$ brings the Bethe equations (\ref{xxzbethe}) to the form:
\begin{equation}
e^{i (M+1-N) \theta_k}=(-1)^{N-1}\prod_{j=1}^N e^{-i\theta_j}\,, \quad 1 \le k\le N\,,
\label{bthv}
\end{equation}
which is respected by the solutions
\begin{equation}
\theta_j=\frac{2\pi}{M+1-N}\Bigl(I_j-\frac{N-1}{2}-P\Bigr),
\label{sol}
\end{equation}
where $P \equiv \displaystyle{\frac{1}{2\pi}\sum_{j=1}^N \theta_j}$, and the $I_j$ are integers or half-integers, depending on whether $N$ is odd or even. It suffices to consider a set of $N$ different numbers $I_j$ satisfying the condition  $M-N\geq I_1 > I_2 > \dots> I_N\geq 0$.
The ground state of the model is defined by the solutions
\begin{equation}
\ta^{\rm v}_j\equiv\frac{2\pi}{M+1-N} \Bigl(N-j-\frac{N-1}{2}\Bigr),
\label{grstsa}
\end{equation}
with the condition $P=0$ fulfilled. The notation $\bth$ for $N$-tuple $(\ta_{1}, \ta_{2}, \dots, \ta_{N})$ of the solutions (\ref{sol}) will be especially convenient further in order to stress that one is concerned with the solution of the Bethe equations. Otherwise, it is appropriate to use ${\textbf u}$ as an indication that an arbitrary parametrization of the wave-function (\ref{xxwf1}) is meant. The eigen-energy of the ground state of the model calculated by means of (\ref{een}) and (\ref{grstsa}) is:
\begin{equation}
E_N(\bth^{\rm v})\,=\,
h N\,-\,{\text{cosec}}\frac{\pi}{M+1-N} \,\sin\frac{\pi N}{M+1-N} \,.
\label{eigenv}
\end{equation}
The eigen-energy of the Ising Hamiltonian (\ref{ising}) has the form:
$E_{\rm Is} = N$.

The two limits, (\ref{anis1}) and (\ref{anis2}), are similar in the sense that their wave functions are expressible through the {\it Schur functions} \cite{macd}:
\begin{equation}
S_{\bla} ({\textbf x})\,\equiv\,
\displaystyle{
S_{\bla} (x_1, x_2, \dots, x_N) \,\equiv\,\frac{\det(x_j^{\la_k+N-k})_{1\leq j,k\leq
N}}{\det(x_j^{N-k})_{1\leq j, k\leq N}}}  \,,
\label{sch}
\end{equation}
where ${\bla}$ denotes the partition $(\la_1, \la_2, \dots, \la_N)$, which is an $N$-tuple of nonincreasing non-negative integers:
$L\geq\la_1\ge\la_2\ge\,\dots\,\ge \la_N\geq 0$. Indeed, any strict partition $M\geq\mu_1>\mu_2>\,\dots\,> \mu_N\geq 0$ and non-strict partition $M+1-N\geq \la_1\geq \la_2\geq \dots\geq \la_N\geq 0$ (denoted as $\bmu$ and $\bla$, respectively) can be related by the formula $\la_j=\mu_j-N+j$, where $1\le j\le N$.
In other terms, $\bla =\bmu -\bdl$, where $\bdl$ is the strict partition $(N-1, N-2, \dots, 1, 0)$. Any strict partition $\bmu$ with the elements respecting the exclusion condition $\mu_i>\mu_{i+1}+1$ is connected with the  non-strict partition $\widetilde{\bla}$ by the relation $\widetilde{\bla} =\bmu -2\bdl$, where $M+2(1-N)\geq \widetilde \la_1 \geq \widetilde\la_2\geq \dots\geq \widetilde \la_N\geq 0$. So, the wave function (\ref{xxwf1}) may be represented as
$
\chi_{\bmu}^{\rm SA} ({\textbf u})=S_{\widetilde{\bla}}
({\textbf u}^2)
$,
and the corresponding state-vector (\ref{bwf}) is specialized as follows:
\begin{equation}
\mid\!\Psi_N({\textbf u})\rangle = \sum_{\widetilde\bla \subseteq \{(M-2(N-1))^N\}} S_{\widetilde\bla}
({\textbf u}^2)\prod\limits_{k=0}^M (\si_k^{-})^{e_k}\mid\Uparrow\rangle\,.
\label{vstv}
\end{equation}
Summation in (\ref{vstv}) goes over all non-strict partitions $\widetilde{\bla}$ into at most $N$ parts so that each is less than $M-2(N-1)$, i.e., $M-2(N-1)\ge\widetilde \la_1 \geq \widetilde\la_2\geq \dots\geq \widetilde \la_N\geq 0$. Summation over the non-strict partitions $\widetilde{\bla}$ is equivalent to that over the strict partitions $\bmu$ with the elements respecting the conditions $\mu_i>\mu_{i+1}+1$ and $\widetilde{\bla}=\bmu -2\bdl$.

\section{Survival probability of the ferromagnetic string}\label{tmf:sec3}

We shall study the thermal correlation function of the states with no spins ``down'' on the last $n+1$ sites of the lattice (would be called as {\textit {survival probability of the ferromagnetic string}}) defined by the ratio:
\begin{equation}
\CT ({\bth}^{\rm v}, n, \be)\,\equiv\,\frac{\langle
\Psi_N({\bth}^{\rm v})\mid \bar\varPi_{n}\,e^{-\be \widehat H_{\rm SA}}\,
\bar\varPi_{n} \mid\!\Psi_N({\bth}^{\rm v})\rangle }{\langle
\Psi_N({\bth}^{\rm v})\mid \,e^{-\be \widehat H_{\rm SA}}\,
\mid\!\Psi_N({\bth}^{\rm v})\rangle}\,,\qquad \bar\varPi_{n} \equiv \prod\limits_{j=M-n}^M \check q_j\,,
\label{ratbe0}
\end{equation}
where $\widehat H_{\rm SA}$ and $\bar\varPi_{n}$ are defined, respectively, by (\ref{anis2}) and (\ref{proj}), and $\be$ is inverse temperature. Besides, $\bth^{\rm v}$ indicates that the eigen-states are calculated on the solution (\ref{grstsa}). Our calc\-ulat\-ions will be based on the following relation for the Schur functions (\ref{sch}) which is due to the Binet--Cauchy formula \cite{gant}:
\begin{eqnarray}
\sum_{\bla \subseteq \{L^N\}}S_{\bla} (x_1^2, \dots,
x_N^2) S_{\bla} (y_1^2, \dots, y_N^2)\,=\,\det \Bigl(\frac{1-(x_k y_j)^{2(N+L)}}{1-(x_k y_j)^2}\Bigr)_{1\le j, k\le N}   \nonumber \\
  \times \prod_{1\leq k<j \leq N} \left( y_j^2-y_k^2\right)^{-1} \prod_{1\leq m<l\leq N} \left(x_l^2-x_m^2\right)^{-1}\,.  \label{schr}
\end{eqnarray}
Summation in (\ref{schr}) goes over all non-strict partitions $\bla$ into at most $N$ parts so that each is less than $L$: $L\ge \la_1 \ge \la_2\ge \dots \ge \la_N \ge 0$. The notation for the range of summation in (\ref{schr}) will be extensively used in the rest of the paper.

Firstly,
it is straightforward to calculate, using (\ref{schr}), the following average of the projector $\bar\varPi_{n}$ (\ref{anis2}) over the state-vectors (\ref{vstv}):
\begin{eqnarray}
\langle \Psi_N({\textbf v})\mid\bar\varPi_{n}\mid\!\Psi_N({\textbf u})\rangle\,=\,\sum_{\widetilde{\bla} \subseteq \{(M-2N-n+1)^N\}}S_{\widetilde {\bla}} ({\textbf v}^{-2}) S_{\widetilde {\bla}} ({\textbf u}^2) &&  \nonumber \\
=\displaystyle{ \frac{1}{{\CV}({\textbf u}^2){\CV}({\textbf
v}^{-2})} \det\Bigl(\frac{1 - (u_k^2/v_j^2)^{M-N-n+1}}{1 -
u_k^2/v_j^2}\Bigr)_{1\le j, k \le N}}\,,&&
\label{spdfpxx}
\end{eqnarray}
where the notation for the Vandermonde determinant is used:
\begin{equation}
\CV({\textbf u}^{2})\,\equiv\,\prod_{1\leq m<l\leq N}(u_l^2-u_m^2)\,. \label{spxx1}
\end{equation}
Right-hand side of (\ref{spdfpxx}), taken at $n+1=0$, gives the answer for the scalar product of the state-vectors (\ref{vstv}), i.e., for $\langle \Psi_N({\textbf v})\mid\!\Psi_N({\textbf u})\rangle$. Equation (\ref{spdfpxx}), taken on the solutions (\ref{grstsa}), allows one to obtain the so-called \textit{Emptiness Formation Probability}  $E(n, N)$, which expresses the probability of formation of a string of the empty (i.e., spin ``up'') states on last $n+1$ sites of the lattice:
\begin{eqnarray}
E(n, N)\,\equiv\,\frac{\langle
\Psi_N({\bth}^{\rm v})\mid \bar\varPi_{n}\, \mid\!\Psi_N({\bth}^{\rm v})\rangle }{\langle
\Psi_N({\bth}^{\rm v}) \mid\!\Psi_N({\bth}^{\rm v})\rangle}=\,\frac{M-N+1}{M+1}\,\det \Bigl(\bigl(1-\frac{n}{M-N+1}\bigr)\dl_{lj} \nonumber \\ [0.2cm] +\,\frac{1-e^{i n(k_l-k_j)}
}{(M-N+1)(1-e^{i(k_l-k_j)})} (1-\delta_{l j})\Bigr)_{1\leq l, j\leq N}\,,\qquad k_l\equiv \frac{2\pi l}{M+1-N}.&& \nonumber
\end{eqnarray}

The solutions of the Bethe equations (\ref{bthv}) constitute a complete set of the eigen-states, and therefore one can obtain the resolution of the identity operator:
\begin{equation}
{\BI}\,=\,\frac{1}{(M+1)(M+1-N)^{N-1}}
\sum\limits_{\{{\bth}\}}\,\bigl| {\CV}(e^{i{\bth}})  \bigr|^2
\mid\!\Psi_N({\bth})\rangle \langle \Psi_N({\bth})\!\mid\,,
\label{ratbe171}
\end{equation}
where the summation over $\{{\bth}\}$ implies summation over all independent solutions (\ref{sol}). Here the exponential parametrization for the solutions of the Bethe equations is used in the compact form,
$
e^{i \bth} \equiv (e^{i \ta_{1}}, e^{i \ta_{2}}, \dots, e^{i \ta_{N}})
$.
We shall calculate the nominator of (\ref{ratbe0}) using insertions of the resolution (\ref{ratbe171}). Taking into account that
\[\langle \Psi_N({\textbf v})\mid e^{-\be \widehat H_{\rm SA}}\mid\!\Psi_N(\bth)\rangle\,=\,\langle \Psi_N({\textbf v})\mid\!\Psi_N({\bth})\rangle\,e^{-\be E_N (\bth) },
\]
we obtain:
\begin{equation}
\begin{array}{r}
\displaystyle{\langle \Psi_N({\textbf v})\mid \bar\varPi_{n}\, e^{-\be {\widehat H}_{\rm SA}}\,
\bar\varPi_{n} \mid\!\Psi_N({\textbf u})
\rangle\,= \,\frac{1}{(M+1)(M+1-N)^{N-1}} \sum\limits_{\{{\bth}\}}
\,e^{-\be E_N(\bth)}}\\[0.6cm]
\times\,\displaystyle{\bigl| {\CV}(e^{i{\bth}})  \bigr|^2\,\CP({\textbf v}^{-2}, e^{i{\bth}})\,\CP(e^{-i{\bth}}, {\textbf u}^{2})}\,,
\end{array}
\label{ratbe200}
\end{equation}
where
\begin{equation}
{\CP}({\textbf v}^{-2},
e^{i\bth})\,\equiv\,\sum\limits_{{\widetilde\bla^L}} S_{{\widetilde\bla^L}}({\textbf
v}^{-2})\, S_{{\widetilde\bla^L}}(e^{i\bth})\,,\quad {\CP}(e^{-i\bth},
{\textbf u}^{2})\,\equiv\, \sum\limits_{{\widetilde\bla^R}}
S_{{\widetilde\bla^R}}(e^{-i\bth})\, S_{{\widetilde\bla^R}}({\textbf u}^2)\,.
\label{ratbe201}
\end{equation}
The range of summation in (\ref{ratbe201}) is taken as follows: ${\widetilde\bla^L},\,{\widetilde\bla^R} \subseteq \{(M-2N-n+1)^N\}$.
Then, using the relation (\ref{schr}) we obtain from (\ref{ratbe200}):
\begin{equation}
\begin{array}{rcl}
&&\langle \Psi_N({\textbf v})\mid \bar\varPi_{n}\, e^{-\be {\widehat H}_{\rm SA}}\,
\bar\varPi_{n} \mid\!\Psi_N({\textbf u})
\rangle\,= \\[0.2cm]
&=&
\displaystyle{\frac{1}{(M+1)(M+1-N)^{N-1}\,{\CV}({\textbf u}^2){\CV}({\textbf v}^{-2})} \sum\limits_{M-N\ge I_1 > I_2 \dots > I_N\ge 0}
\!\!e^{-\be E_N(\bth)}}\\[0.6cm]
&\times&\displaystyle{\det \Bigl(
\frac{1-(e^{i\ta_{i}}v_j^{-2})^{M-N-n+1}}{1- e^{i\ta_{i}}v_j^{-2}}
\Bigr)_{1\le i, j \le N}\,\det
\Bigl(\frac{1-(u_l^2 e^{-i\ta_{p}})^{M-N-n+1}}{1-u_l^2
e^{-i\ta_{p}}}\Bigr)_{1\le l, p \le N}}\,,
\end{array}
\label{ratbe20}
\end{equation}
where summation goes over the ordered sets $\{I_k\}_{1\le k\le N}$ that parametrize $\bth$ (\ref{sol}), and
$E_N(\bth)$ is given by (\ref{eigenv}).
Expression for $\CT ({\bth}^{\rm v}, n, \be)$ (\ref{ratbe0}) appears from (\ref{ratbe20}) as follows:
\begin{equation}
\begin{array}{r}
\CT ({\bth}^{\rm v}, n, \be)
=\displaystyle{\frac{1
}
{(M+1)^2(M+1-N)^{N-2}}\!\sum\limits_{{M-N}\ge I_1
> I_2 \dots > I_N\ge 0}
\!\!e^{-\be (E_N(\bth)-E_N(\bth^{\rm v}))}}  \\[0.3cm]
\times\,\displaystyle\Biggl|{\det
\Bigl(\frac{1-e^{i(M-N-n+1)(\ta_l-{\ta}^{\rm v}_{p})}}{1-
e^{i(\ta_l-{\ta}^{\rm v}_{p})}}\Bigr)_{1\le p, l \le N}\Biggr|^{2}},
\end{array}
\label{ratbe212}
\end{equation}
where $\bth^{\rm v}$ implies the ground-state solution (\ref{grstsa}), and $E_N(\bth^{\rm v})$ is the ground-state energy in the limit of strong anisotropy.

\section{Four-vertex model and boxed plane partitions}\label{tmf:sec4}

\subsection{The $\rm\bf XXZ$ Hamiltonian at $\Dl\to-\infty$ and the four-vertex model}
The \textit{six-vertex} model on a square lattice is defined by six
different configurations of arrows pointed both in and out of each
lattice site. A statistical weight $w_k$ $(k=1, 2, \dots , 6)$ is
ascribed to each admissible type of the vertices (Fig. 1).
Representing the arrows pointing up or to the right by the solid
lines one can get the alternative description of the vertices in
terms of lines floating through the lattice sites. So far the
bonds of a lattice may be only in two states -- with a line either
without it, there is a one to one correspondence  between the
admissible configuration of arrows on a lattice and the network of
lines -- the nest of lattice paths. The \textit{four-vertex} model is obtained in the limit when the weights
$w_1=w_3=0$.

\begin{figure}[h]
\centering
\includegraphics{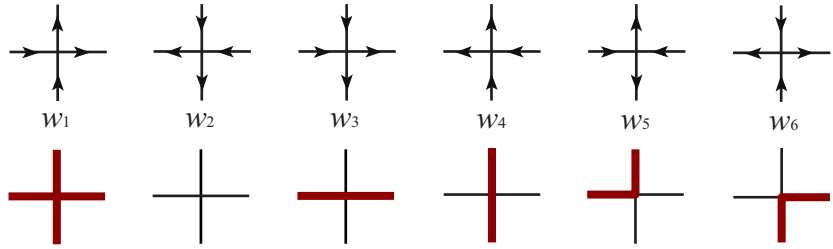}
\caption{
The vertex configurations of the 6-vertex model.}
\end{figure}

The $L$-operator of the six-vertex model is equal to \cite{ft}:
\begin{equation}
L_{\rm 6v}(n|u)\,=\,\Biggl(
\begin{array}{cc}
-ue^{\ga\si_n^z}-u^{-1}e^{-\ga\si_n^z} & 2\sinh(2\ga)\, \si_n^{-} \\ 2\sinh(2\ga)\,
\si_n^{+}  &
ue^{-\ga\si_n^z}+u^{-1}e^{\ga\si_n^z}
\end{array}
\Biggr)\,,
\label{ratbe50}
\end{equation}
where $u\in\BC$ and $\ga\equiv\frac12{\rm Arch} \Dl$,
and it satisfies the intertwining relation:
\begin{equation}
\widetilde R(u,v)\bigl(L_{\rm 6v}(n|u)\,\otimes\,L_{\rm 6v}(n|v)\bigr)\,=\,\bigl(L_{\rm 6v}(n|v)\,\otimes\,L_{\rm 6v}(n|u)\bigr)\widetilde R(u,v)\,.
\label{rsv0}
\end{equation}
Here, $\widetilde R(u,v)$ is the $(4\times 4)$-matrix:
\begin{equation}
\widetilde R(u,v)=\left(
\begin{array}{cccc}
\widetilde f(v,u) & 0 & 0 & 0 \\
0 & \widetilde g(v,u) & 1 & 0 \\
0 & 1 & \widetilde g(v,u) & 0 \\
0 & 0 & 0 & \widetilde f(v,u)
\end{array}
\right) ,  \label{rsv1}
\end{equation}
where
\begin{equation}
\widetilde f(v,u)=\frac{u^2e^{2\ga }-v^2e^{-2\ga
}}{u^2-v^2}\,,\,\qquad\widetilde g(v,u)=\frac{uv}{u^2-v^2}\left( e^{2\ga
}-e^{-2\ga }\right)\,,
\label{fgsv}
\end{equation}
and $u, v \in\BC$.

Let us consider the following transformation of the $L$-operator
(\ref{ratbe50}):
\begin{eqnarray}
\Check L(n|u) &=&e^{h\,\si_n^z}e^{(\omega/ 2)\,\si^z} L_{\rm 6v}(n|u)\,e^{-(\omega/ 2)\,\si^z}  \label{tlsv} \\[0.3cm]
&=&\left(
\begin{array}{cc}
-ue^{(h+\gamma )\si_n^z}-u^{-1}e^{(h-\ga )\si_n^z} & 2\sinh(2\ga)\,
e^{\omega+h\si_n^z}\,\si_n^{-}  \\ 2\sinh(2\ga)\,
e^{-\omega +h\si_n^z}\,\si_n^{+}  & u e^{(h-\ga )\si_n^z}+u^{-1}e^{(h+\ga )\si_n^z}
\end{array}
\right) .  \nonumber
\end{eqnarray}
The operator (\ref{tlsv}) is intertwined by the transformed $R$-matrix of the form:
\begin{eqnarray}
\check R(u,v) &=&\bigl( 1\otimes e^{-h\si^z}\bigr) \widetilde
R(u,v)\bigl(
1\otimes e^{h \si^z}\bigr)   \label{trsv} \\
&=&\left(
\begin{array}{cccc}
\widetilde f(v,u) & 0 & 0 & 0 \\
0 & \widetilde g(v,u) & e^{2h} & 0 \\
0 & e^{-2h} & \widetilde g(v,u) & 0 \\
0 & 0 & 0 & \widetilde f(v,u)
\end{array}
\right) .  \nonumber
\end{eqnarray}
If one puts  $h=\omega =\ga$, then the limit
\begin{equation}
\lim_{\gamma \rightarrow\infty} e^{-2\gamma }\check L(n|u)=L_{\rm 4v}(n|u)\,\equiv\,\left(
\begin{array}{cc}
-u\check q_n & \si_n^{-} \\
\si_n^{+} & u^{-1} \check q_n
\end{array}
\right)\,,  \label{ratbe51}
\end{equation}
where $\check q_n =\si^+_n \si^-_n$ (see (\ref{proj})), gives us the $L$-operator of, so-called, {\it four-vertex} model.
Analogously, the limit $\ga\to\infty$
transforms the matrix $e^{-2\ga} \check R(u,v)$ into the $R$-matrix of the four-vertex model \cite{b7}:
\begin{equation}
R(u,v)=\left(
\begin{array}{cccc}
f(v,u) & 0 & 0 & 0 \\
0 & g(v,u) & 1 & 0 \\
0 & 0 & g(v,u) & 0 \\
0 & 0 & 0 & f(v,u)
\end{array}
\right) ,  \label{r}
\end{equation}
where
\begin{equation}
f(v,u)=\frac{u^2}{u^2-v^2}\,,\qquad g(v,u)=\frac{uv}{u^2-v^2}\,.  \label{fg}
\end{equation}

The monodromy matrix of the models is defined as the matrix product of $L$-operators:
\begin{equation}
T(u)=L(M|u)L(M-1|u)\,\cdots\,L(0|u)=\Biggl(
\begin{array}{cc}
A(u) & B(u) \\
C(u) & D(u)
\end{array}
\Biggr).  \label{ratbe52}
\end{equation}
The transfer matrix is the matrix trace of the monodromy matrix: $\tau(u)\,=\,{\rm Tr}\,T(u)$.
In \cite{lib} it was proved that the transfer-matrix of the six-vertex model commutes with the Hamiltonian of the $XXZ$ model (\ref{xxzham}): $[{\widehat H}_{\rm 6v},\tau_{\rm 6v}(u)]=0$.

The transfer-matrix of the four-vertex model $\tau_{\rm 4v}(u)$ satisfies the property:
\begin{equation}
\mathcal{P} \tau_{\rm 4v}(u)= \tau_{\rm 4v}(u)
\mathcal{P}=\tau_{\rm 4v}(u),
\label{lt}
\end{equation}
where $\mathcal{P}$ is the projector defined in (\ref{anis2}). Indeed, using the explicit expression for the entries of the product of two $L$-operators,
\begin{equation}
L(n+1|u)L(n|u)=\left(
\begin{array}{cc}
A_{n+1,n} & B_{n+1,n} \\
C_{n+1,n} & D_{n+1,n}
\end{array}
\right)\,,
\label{ratbe543}
\end{equation}
one can show that
\begin{equation}
\CC_{n+1, n}\,L(n+1|u)L(n|u)=L(n+1|u)L(n|u)\,\CC_{n+1, n}=L(n+1|u)L(n|u)\,,
\label{cl}
\end{equation}
where $\CC_{n+1, n}\equiv 1-\hat q_{n+1}\hat q_n$. It follows from (\ref{cl}) that $\CC_{n+1, n}\,\tau_{\rm 4v}(u)=\tau_{\rm 4v}(u)\,\CC_{n+1, n}=\tau_{\rm 4v}(u)$ at $n=0,\ldots,M-1$. Invariance of the matrix trace with respect of cyclic permutation implies that $\CC_{0, M}\,\tau_{\rm 4v}(u)=\tau_{\rm 4v}(u)\,\CC_{0, M}=\tau_{\rm 4v}(u)$, and therefore the relation (\ref{lt}) is also fulfilled.

Further, one can advance the following

{\bf Proposition:}
{\it The transfer matrix of the four-vertex model commutes with the $XXZ$
Hamiltonian taken in the Ising limit}:
\begin{equation}
[{\widehat H}_{\rm SA},\tau_{\rm 4v}(u)]=0\,.\label{ratbe53}
\end{equation}

{\bf Proof:} In order to prove (\ref{ratbe53}), it is suffices to consider ${\widehat H}_{\rm SA}$ (\ref{anis2}) at $h=0$. We shall use the method proposed in \cite{lib} and then developed in
\cite{tar}. Generalizing this approach with respect of the problem under consideration, we shall take into account the relation (subscripts are omitted):
\begin{equation}
\begin{array}{l}
\CC_{n+1, n}\,\bigl[\widehat h_{n+1,n},L(n+1|u)L(n|u)\bigr]\,\CC_{n+1, n} \\[0.3cm]
=\,\CC_{n+1, n}\bigl( Q(n+1|u)L(n|u)-L(n+1|u)Q(n|u)\bigr)
\CC_{n+1, n}\,,
\end{array}
\label{ratbe54}
\end{equation}
where the operators $\widehat h_{n+1,n}$ and $\CC_{n+1, n}$ are introduced in (\ref{anis1}) and (\ref{cl}), respectively. If so, we obtain the vanishing of the commutator in question,
\begin{equation}
\bigl[{\widehat H},\tau (u)\bigr] \,=\,-\frac 12\sum_{n=0}^M \CP\,\bigl[\widehat h_{n+1,n},\tau
(u)\bigr]\,\CP\,=\,-\frac
12 \sum_{n=0}^M \Tr\bigl(\CP\,\bigl[ \widehat h_{n+1,n},T(u)\bigr]\,\CP\bigr)\,=\,0\,,
\label{ratbe541}
\end{equation}
provided that
\begin{equation}
\bigl[\widehat h_{n+1,n},T(u)\bigr]\,=\,
L(M|u) \cdots L(n+2|u)\,\bigl[
\widehat h_{n+1,n},L(n+1|u)L(n|u) \bigr]\,L(n-1|u)\cdots L(0|u)\,.\label{ratbe542}
\end{equation}

Furthermore, the commutators of the entries from the right-hand side of (\ref{ratbe543}) look as follows:
\begin{equation}
\begin{array}{l}
\bigl[\widehat h_{n+1,n},A_{n+1,n}\bigr]\,=\,-\,\bigl [\widehat h_{n+1,n},D_{n+1,n} \bigr ]\,=\,\check q_{n+1}-\check q_n \\[0.3cm]
\bigl [\widehat h_{n+1,n},B_{n+1,n}\bigr] \,=\,-u\si_{n+1}^{-}+u^{-1}
\si_n^{-}+u\si_{n+1}^{-} \hat q_n-u^{-1} \hat q_{n+1}
\si_n^{-}, \\[0.3cm]
\bigl [\widehat h_{n+1,n},C_{n+1,n} \bigr] \,=\,-u^{-1}\si_{n+1}^{+}+u
\si_n^{+}+u^{-1}\si_{n+1}^{+} \hat q_n-u \hat q_{n+1}
\si_n^{+}\,.
\end{array}
\label{ratbe544}
\end{equation}
In turn, one obtains from (\ref{ratbe544}):
\begin{equation}
\begin{array}{l}
\CC_{n+1, n}\, \bigl[\widehat h_{n+1,n}, A_{n+1,n}\bigr]\,\CC_{n+1, n}
\,=\,\CC_{n+1, n}\bigl (
\check q_{n+1}-\check q_n\bigr ) \CC_{n+1, n}, \\[0.3cm]
\CC_{n+1, n}\,\bigl[\widehat h_{n+1,n}, B_{n+1,n}\bigr]\,\CC_{n+1, n}
\,=\,\CC_{n+1, n} \bigl(
-u\si_{n+1}^{-}+u^{-1}\si_n^{-}\bigr ) \CC_{n+1, n}, \\[0.3cm]
\CC_{n+1, n}\,\bigl[\widehat h_{n+1,n}, C_{n+1,n}\bigr]\,\CC_{n+1, n}
\,=\,\CC_{n+1, n}\bigl(
-u^{-1}\si_{n+1}^{+}+u\si_n^{+}\bigr ) \CC_{n+1, n}, \\[0.3cm]
\CC_{n+1, n}\,\bigl[\widehat h_{n+1,n}, D_{n+1,n}\bigr]\,\CC_{n+1, n}
\,=\,\CC_{n+1, n}\bigl( -\check q_{n+1} + \check q_n \bigr ) \CC_{n+1, n}.
\end{array}
\label{ratbe545}
\end{equation}
On the other hand, using the diagonal matrix
\[
Q(n|u)=\left(
\begin{array}{cc}
u^{-1} \si^0 & 0 \\
0 & u \si^0
\end{array}
\right)\,,
\]
we obtain the following relation for the $L$-operators (\ref{ratbe51}):
\begin{equation}
Q(n+1|u)L(n|u)-L(n+1|u)Q(n|u)=\left(
\begin{array}{cc}
\check q_{n+1}-\check q_n & -u\si_{n+1}^{-}+u^{-1}\si_n^{-} \\
-u^{-1}\si_{n+1}^{+}+u\si_n^{+} & -\check q_{n+1}+\check q_n
\end{array}
\right)\,.
\label{ratbe546}
\end{equation}
Comparison of (\ref{ratbe545}) and (\ref{ratbe546}) demonstrates that the relation (\ref{ratbe54}), as well as the statement of the relation~(\ref{ratbe53}) are valid indeed. $\quad\blacksquare$

The state vector of the four-vertex model is constructed in the framework of the algebraic Bethe ansatz as follows:
\begin{equation}
\bigl| \Psi _N(\textbf {u})\rangle=\prod_{i=1}^N B_{\rm 4v}(u_i) \,\bigl| \Uparrow \rangle\,.
\label{stva}
\end{equation}
This vector is the eigen-vector both of the  transfer-matrix of the four-vertex model and of the Hamiltonian (\ref{anis2}), provided that the parameters $u_l$ fulfill the Bethe equations (\ref{bthv}).
It was shown in \cite{b7} that this vector can be represented in the form (\ref{vstv}).

According to the representation (\ref{spdfpxx}) taken at $n+1=0$, the scalar product \break $\langle \Psi_N({\textbf v})\mid\!\Psi_N({\textbf u})\rangle$ is proportional to the determinant of the matrix, say, $(T_{k j})_{1\leq k, j\leq N}$. The Bethe equations (\ref{bthv}) enable one to represent this determinant
as follows:
\begin{equation}
\det(T_{k j})_{1\leq k, j\leq N}\,
\equiv\,\det\Biggl(\frac{1-(u^2_k/v^2_j)^{M-N+2}}{1-u^2_k/v^2_j}\Biggr)
=\,\Biggl(\prod\limits_{k=1}^N\frac{-v_k}{u_k}\Biggr)^N \det(A_{k j})_{1\leq k, j\leq N}
 \,,
\label{normv1}
\end{equation}
where the entries $A_{k j}$ take the form:
\begin{equation}
\displaystyle{
A_{k j}\,=\,\frac{(u_k v_j)^{M+3-N}}{u_k^2- v_j^2}\,\Biggl(\prod\limits_{l=1,\,l\ne k}^N u_l^2\,-\,\!\!\!\prod\limits_{l=1,\,l\ne j}^N v_l^2\Biggr)}
 \,.
\label{normv2}
\end{equation}
It is crucial that the determinant of the matrix $(A_{k j})_{1\leq k, j\leq N}$ vanishes provided the par\-amet\-ers ${\textbf u}$ and ${\textbf v}$ are independent solutions of the Bethe equations (\ref{bthv}). Really, this matrix has a non-trivial eigen-vector $\chi_j$ with zero eigen-value:
\begin{equation}
\displaystyle{
\sum\limits_{j=1}^N A_{k j} \chi_j\,=\,0\,,\qquad \chi_j\,=\,(u_j^2- v_j^2)\prod\limits_{l=1,\,l\ne j}^N \frac{u_j^2- v_l^2}{u_j^2- u_l^2}}
 \,.
\label{normv3}
\end{equation}
Validity of (\ref{normv3}) is justified by two identities:
\[
\begin{array}{l}
\displaystyle{
\sum\limits_{k=1}^N \Biggl(\frac{y_k-x_k}{y_k-x_j}
\prod\limits_{l=1,\,l\ne k}^N \frac{y_k-x_l}{y_k-y_l}\Biggr)\,=\,1
}\,,\\[0.6cm]
\displaystyle{\sum\limits_{k=1}^N\Biggl(\frac{y_k-x_k}{y_k}
\prod\limits_{l=1,\,l\ne k}^N \frac{y_k-x_l}{y_k-y_l}\Biggr)\,=\,1\,-\,\prod\limits_{l=1}^N \frac{x_l}{y_l}}\,.
\end{array}
\]
Therefore, the scalar product $\langle
\Psi_N({\textbf v})\!\mid\!\Psi_N({\textbf u}) \rangle$ vanishes (i.e., the state-vectors are or\-tho\-go\-nal) provided the sets of the parameters ${\textbf u}$ and ${\textbf v}$ represent independent Bethe solutions. Completeness of the system of the state-vectors was proved in \cite{god, abar}.

\subsection{Boxed plane partitions}

Consider the four-vertex model \cite{b7}. There is one to one cor\-res\-pon\-dence between the strict plane partitions (i.e., the plane partitions that decay along each column and each raw \cite{macd}) and the nests of the admissible paths on a square lattice of the size $2N\times (M+1)$ with the following boundary conditions: all arrows on the left and
right boundaries are pointing to the left, while the arrows on the
top and bottom of the first $N$ columns (counting from the left)
are pointing inwards and the arrows on the top and bottom of the
last $N$ ones are pointing outwards (see Fig. 2).
\begin{figure}[h]
\centering
\includegraphics{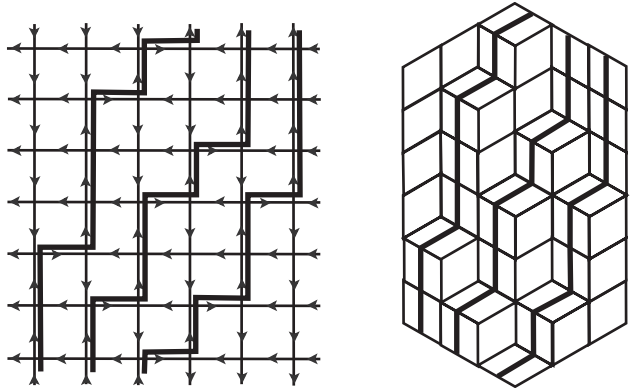}
\caption{The admissible nest of lattice paths of the four-vertex model and the corresponding strict plane partition.}
\end{figure}

It was proved in \cite{b7} that if one puts  $v_j=q^{-\frac j2}$ and
$u_j=q^{\frac{j-1}2}$ in the scalar product of the state-vectors (\ref{stva}), then
\begin{equation}
\langle \Psi_N(q^{-\frac 12}, \dots, q^{-\frac N2})\mid\!\Psi_N
(1, \dots, q^{\frac{N-1}2})\rangle\,=\, q^{-N^2 (N-1)}\,Z^{\rm spp}_q(N,N,M)\,,
\label{ratbe57}
\end{equation}
where $Z^{\rm spp}_q(N,N,M)$ is the generating function
of strict plane partitions placed into a box of size $N\times N\times
M$:
\begin{eqnarray}
Z^{\rm spp}_q(N,N,M)\,\equiv\,q^{N^2 (N-1)}\,\prod_{1\leq j,k \leq
N}\frac{1-q^{M+3-j-k}}{1-q^{j+k-1}}.  \label{ratbe58}
\end{eqnarray}
Being taken at $q=1$, this formula gives the number of strict plane
partitions in a box of size $N\times N\times M$:
\begin{eqnarray}
 Z^{\rm spp}_{q=1}(N,N,M)=\prod_{1\leq j,k \leq N}\frac{M+3-j-k}{j+k-1}.
\label{ratbe59}
\end{eqnarray}

It is straightforward to obtain \cite{b8,b5} that the entry
\begin{equation}
\lim_{q\to 1}\,\langle \Psi_N (q^{-\frac 12}, \dots, q^{-\frac N2})\mid\bar\varPi_{n}\mid\!\Psi_N(1, \dots, q^{\frac{N-1}2})\rangle\,=\,
Z^{\rm spp}_{q=1}(N,N,M-n-1)\,
\label{spdfpxx3}
\end{equation}
is equal to  the number strict  partitions in a $N\times N\times (M-n-1)$ box.

\section{Low temperature limit}\label{tmf:sec5}

Let us obtain the low temperature estimate of the correlation function (\ref{ratbe0}). The survival probability of the ferromagnetic string is written as follows:
\begin{equation}
\CT ({\bth}^{\rm v}, n, \be)\,=\,\frac{\bigl|{\CV}(e^{i {\bth}^{\rm v}})\bigr|^2}{\,{\sf V}^2}\,
\displaystyle{\sum\limits_{\{\bth\}}
e^{-\be (E_N(\bth)-E_N(\bth^{\rm v}))}}\,
\bigl|\displaystyle{{\CV}(e^{i{\bth}})\, {\CP}(e^{-i\bth}, e^{i {\bth}^{\rm v}})\bigr|^2\,,}
\label{ratbe61}
\end{equation}
where ${\sf V}\equiv(M+1)(M+1-N)^{N-1}$.
Here ${\CP}(e^{-i\bth}, e^{i {\bth}^{\rm v}})$ is given according to (\ref{ratbe201}), provided the summation domain is given by ${\widetilde\bla} \subseteq \{(M-2N-n+1)^N\}$.

If the chain is long enough while the number of quasi-particles
is moderate, i.e., $N \ll M$, we replace  the sums in (\ref{ratbe61}) by the integrals as follows:
\begin{equation}
\begin{array}{r}
\CT ({\bth}^{\rm v}, n, \be)\,\simeq\,\displaystyle{\frac{1}{(2\pi)^N N!}\,\prod_{1\leq r<s\leq N}
2\bigl(1-\cos({\ta}^{\rm v}_r-{\ta}^{\rm v}_s)\bigr)\, }\\[0.6cm]
\displaystyle{\times\,\prod\limits_{i=1}^N\Bigl(
\int_{0}^{2\pi}\!\frac{d\ta_i}{2\pi}\Bigr)
\,e^{{\be}\sum\limits_{l=1}^N(\cos{\ta}_l-\cos{\ta}^{\rm v}_{l})}\,\bigr|{\CP}( e^{-i{\bth}}, e^{i {\bth}^{\rm v}})\bigr|^2}\prod_{1\leq k<l\leq N} \bigl|e^{i{\ta}_k}-e^{i{\ta}_l}\bigr|^2\,.
\end{array}
\label{ratbe62}
\end{equation}
In the considered limit, we may asymptotically put $\cos{\ta}_{l}\approx 1$ and $\cos{\ta}^{\rm v}_{l}\approx 1$, $\forall l$. At the large enough $\be$, we approximate (\ref{ratbe62}) in the following way:
\begin{eqnarray}
\CT ({\bth}^{\rm v}\approx {\textbf 0}, n, \be)\simeq
\displaystyle{\frac{\CI \,{\CP}^2({\textbf 1}, {\textbf 1})}{(2\pi)^N N!}\,\Bigl( \frac{2}{\be}\Bigr)^{N^2/2}\!\!  \prod_{1\leq r<s\leq N}
(r-s)^2} \,,\label{ratbe620}\\[0.3cm]
\displaystyle{\CI\,\equiv\,
\int\limits_{-\infty}^{\infty}
\!\int\limits_{-\infty}^{\infty}
\cdots\int\limits_{-\infty}^{\infty}\,e^{-\sum\limits_{l=1}^N
x^2_l}\!\!\prod_{1\leq k<l\leq N} \bigl|x_k- x_l\bigr|^2
d x_1 d x_2 \dots d x_N
} \,.
\label{ratbe621}
\end{eqnarray}
The integral $\CI$ (\ref{ratbe621}) is the {\textit{Mehta integral}}, and its value is known \cite{meh}:
\[
\CI\,=\,\frac{\pi^{N/2}\,\Ga(1|N+1)}{2^{N(N-1)/2}}\,,
\]
where
\[
\prod_{1\leq r<s\leq N} |r-s| \,=\, \Ga(1|N)\,,\qquad \Ga(1|N)\,\equiv\,\prod\limits_{n=1}^{N} \Ga(n)\,.
\]
Eventually, taking into account (\ref{spdfpxx3}) and (\ref{ratbe620}),
we express the answer in the following form:
\[
\CT_{\rm SA} ({\bth}, n, \be)\,\simeq\,\bigl(\displaystyle{Z^{\rm spp}_{q=1}(N,N,M-n-1)
}\bigr)^2\,\frac{{\cal C}_N}{\be^{N^2/2}}\,,\qquad\,{\cal C}_N\,\equiv\,\frac{
\Ga^3(1|N)}{(2\pi)^{N/2}}\,.
\]
The low temperature decay of the  survival probability of the ferromagnetic string is governed by the critical exponent $N^2/2$, while its amplitude is proportional to the squared number of strict plane partitions in a box of size $N\times
N\times(M-n-1)$.

\section{Conclusion}\label{tmf:sec7}

In this paper we studied the $XXZ$ Heisenberg magnetic chain in a specific limit of the anisotropy parameter: $\Dl\to-\infty$.
The, so-called, survival probability of the ferromagnetic string has been calculated over the $N$-particle ground state of the model.
It was proved that the Hamiltonian of the Ising limit, previously introduced in the papers \cite{alc, abar, b5}, is correct as the limit $\Dl\to-\infty$ of the $XXZ$ Hamiltonian. Finally, we have found the leading term of the low temperature asymptotics of the  survival probability of the ferromagnetic string for the model in the finite volume and at fixed number of quasi-particles. We have demonstrated that the amplitude of the asymptotics is proportional to the squared number of strict plane partitions in a box, while its critical exponent is proportional to the squared number of particles. Comparing this result with that obtained in \cite{b5, bt}, one can conclude that this type of the behavior is universal for a special class of the integrable models.

\section*{Acknowledgement}

This paper was supported in part by the Russian Foundation for
Basic Research \break (No.~10--01--00600) and by the Russian Academy of
Sciences program ``Mathematical Methods in Non-Linear Dynamics''.

\renewcommand{\refname}{

\end {document}
\leftline{{\bf{References}}}}

\begin{thebibliography}{99}

\bibitem{vas1} A. N. Vassiliev, \textit{The Functional Methods in the Quantum Field Theory and Statistics}, Leningrad University Press, Leningrad, 1976. [In Russian]\\
               A. N. Vasiliev, \textit{Functional Methods in Quantum Field Theory and Statistical Physics}, Gordon and Breach, Amsterdam, 1998.

\bibitem{vas2}
    N.~M.~Bogolyubov, V.~F.~Bratsev, A.~N.~Vasil'ev, A.~L.~Korzhenev\-skii, R.~A.~Rad\-zhabov,
    \textit{High temperature expansions at an arbitrary magnetization in the Ising model}, Theor.~Math.~Phys. \textbf{26} (1976), No.~3, 230--237.

\bibitem{yy3}
             C.~N.~Yang, C.~P.~Yang,
             \textit{One-dimensional chain of anisotropic spin-spin interactions. III. Applications}, Phys. Rev., \textbf{151} (1966), No.~1, 258--264.

\bibitem{god}
           M.~Gaudin,
            \textit{La Fonction d'Onde de Bethe}, Masson, Paris, 1983.

\bibitem{f1}
            L.~D.~Faddeev,
            {\it Quantum completely integrable models of field theory},
            Sov. Sci. Rev. Math. C, \textbf{1} (1980), 107--160; In: 40 Years in Mathematical Physics, World Sci. Ser. 20th Century Math., vol.~2, World Sci., Singapore, 1995, pp. 187--235.

\bibitem{KBI1}
            N.~M.~Bogoliubov, A.~G.~Izergin, V. E.~Korepin,
     \textit{Correlation Functions of Inte\-gra\-ble Systems
     and the Quantum Inverse Scattering Method}, Nauka, Moscow,
     1992. [In Russian]

\bibitem{KBI2}
            V.~E.~Korepin, N.~M.~Bogoliubov, A.~G.~Izergin,
     \textit{Quantum Inverse Scattering Method and Correlation
     Functions}, Cambridge University Press, Cambridge, 1993.

\bibitem{vk}
         V.~E.~Korepin,
         \textit{Calculation of norms of Bethe wave functions}, Comm.
          Math. Phys. \textbf{86} (1982), No.~3, 391--418.

\bibitem{vk1}
          A.~G.~Izergin, V.~E.~Korepin,
         \textit{Correlation functions for the Heisenberg $XXZ$-anti\-fer\-ro\-mag\-net}, Comm. Math. Phys. \textbf{99} (1985),  No.~2, 271--302.

\bibitem{ml1}
    N.~Kitanine, J.~M.~Maillet, V.~Terras, \textit{Form factors of the XXZ Heisenberg spin-$\frac12$ finite chain}, Nucl. Phys. B {\bf 554} (1999), No. 3, 647--678.

\bibitem{ml2}
   N.~Kitanine, J.~M.~Maillet, N.~Slavnov, V.~Terras, \textit{Correlation functions of the XXZ spin-$\frac12$ Heisenberg chain at the free fermion point from their multiple integral representations}, Nucl. Phys. B {\bf 642} (2002), No. 3, 433--455.

\bibitem{macd}
            I.~G.~Macdonald,
            {\it Symmetric Functions and Hall Polynomials}, Oxford University Press, Oxford, 1995.

\bibitem{1}
          M. E. Fisher,
          \textit{Walks, walls, wetting, and melting}, J. Statist. Phys., \textbf{34} (1984), No.~5--6, 667--729.

\bibitem{3}
          T. Nagao, P. J. Forrester,
          \textit{Vicious random walkers and a discretization of Gaussian random matrix ensembles}, Nucl. Phys. B, \textbf{620} (2002), No.~3, 551--565.

\bibitem{5}
          C. Krattenthaler, A. J. Guttmann, X. G. Viennot,
          \textit{Vicious walkers, friendly walkers and Young tableaux: II. With a wall}, J. Phys. A: Math. Gen., \textbf{33} (2000), No.~48, 8835--8866.

\bibitem{8}
           M. Katori, H. Tanemura, T. Nagao, N. Komatsuda,
            \textit{Vicious walks with a wall, noncolliding meanders, and chiral and Bogoliubov--de Gennes random matrices},
            Phys. Rev. E, \textbf{68} (2003), No.~2, 021112 [16 pages].

\bibitem{9}
           G. Schehr, S. N. Majumdar, A.~Comtet, J. Randon-Furling,
            \textit{Exact distribution of the maximal height of {\sl p} vicious walkers},
            Phys. Rev. Lett., \textbf{101} (2008), No. 15, 150601 [4 pages].

\bibitem{bres}
            D. M. Bressoud, {\it Proofs and Confirmations. The Story  of the Alternating Sign Matrix Conjecture},
            Cambridge, Cambridge University Press, 1999.

\bibitem{b8}
             N.~M.~Bogoliubov,
           \textit{Boxed plane partitions as an exactely solvable boson model}, J. Phys. A, \textbf{38} (2005), No.~1, 9415-9430.

\bibitem{b1}
          N. M. Bogoliubov,
        \textit{XX Heisenberg chain and random walks}, Zap. Nauchn. Sem. POMI \textbf{325} (2005), 13--27; English transl., J. Math. Sci. \textbf{138} (2006), No.~3, 5636--5643.

\bibitem{b3}
          N.~M.~Bogoliubov, C.~Malyshev,
          \textit{A path integration approach to the correlators of XY Heisenberg magnet and random walks}, In: Proceedings of the 9{\rm th} Intern. Conf. ``Path Integrals: New Trends and Perspectives'' (Dresden, Germany, September 23--28, 2007). Eds., W. Janke, A. Pelster (World Sci., Singapore, 2008), pp. 508--513.
          {\sf arXiv:0810.4816}

\bibitem{b7}
           N.~M.~Bogoliubov,
           \textit{Four-vertex model and random tilings}, Theor. Math. Phys., \textbf{155} (2008), No.~1, 523-535.

\bibitem{b4}
           N.~M.~Bogoliubov, C.~Malyshev,
           \textit{The correlation functions of the XX Heisenberg magnet and random walks of vicious walkers}, Theor. Math. Phys., \textbf{159} (2009), No.~2, 563-574. {\sf arXiv:0903.3227}

\bibitem{b5}
           N.~M.~Bogoliubov, C.~Malyshev,
           \textit{The correlation functions of the $XXZ$ Heisenberg chain in the case of zero or infinite anisotropy, and random walks of vicious walkers}, St. Petersburg Math. J., \textbf{22} (2011), No.~3, 359-377. {\sf arXiv:0912.1138}

\bibitem{lieb}
        E.~Lieb, T.~Schultz, D. Mattis,
        \textit{Two soluble models of an antiferromagnetic chain}, Ann. Phys. (NY), \textbf{16} (1961), No.~3, 407--466.

\bibitem{niem}
         Th. Niemeijer,
         \textit{Some exact calculations on a chain of spin $\frac12$. I, II}, Physica, \textbf{36} (1967), No.~3, 377--419; \textbf{39} (1968), No.~3, 313--326.

\bibitem{iz11}
           F.~Colomo, A.~G.~Izergin, V.~E.~Korepin, V.~Tognetti,
     {\it Temperature correlation functions in the $XXO$ Heisenberg chain}, Theor. Math. Phys., {\bf 94} (1993), No.1, 11--38.

\bibitem{mal1}
        C. Malyshev,
        {\it Functional integration with an ``automorphic'' boundary condition and correlators of third components of spins in the $XX$ Heisenberg model}, Theor. Math. Phys., {\bf 136} (2003), No.~2, 1143--1154.
        {\sf arXiv:hep-th/0204007}

\bibitem{alc}
         F.~C.~Alcaraz, R.~Z.~Bariev,
         \textit{An exactly solvable constrained $XXZ$ chain},
        In: \textit{Statistical Physics on the Eve of the 21$^{st}$ Century. In Honour of J.~B.~McGuire on the Occasion of His 65$^{th}$ Birthday (Series on Advances in Statistical Mechanics)}, Eds., M.~T.~Batchelor, L.~T.~Wille (World Scientific, Singapore, 1999). {\sf arXiv:cond-mat/9904042}

\bibitem{abar}
         N.~I.~Abarenkova, A.~G.~Pronko,
         \textit{Temperature correlation function in the absolutely anisotropic XXZ Heisenberg magnet
}, Theor. Math. Phys., \textbf{131} (2002), No.~2, 690-703.

\bibitem{ess1}
         A. J. A. James, W. D. Goetze, F. H. L. Essler,
         \textit{Finite-temperature dynamical structure factor of the Heisenberg-Ising chain}, Phys. Rev. B, \textbf{79} (2009), No.~21, 214408 [20 pages].

\bibitem{gant}
            F.~R.~Gantmakher,
            {\it Theory of Matrices}, Nauka, Moscow, 1988. [In Russian]

\bibitem{lib}
            E.~H.~Lieb, F.~Y.~Wu,
            \textit{Two dimensional ferroelectric models},
            In: Phase transitions and critical phenomena, vol.~1, Eds., C.~Domb, M.~Green, Academic Press, London, 1972, pp.~331--490.

\bibitem{tar}
         V.~O.~Tarasov, L. A. Takhtadzhyan, L.~D.~Faddeev,
         \textit{Local Hamiltonian for integrable quantum models on a
         lattice}, Theor. Math. Phys. {\bf 57} (1983), No.~2, 1059-1073.

\bibitem{meh}
           M.~L.~Mehta, {\it Random Matrices}, Academic Press, London, 1991.

\bibitem{ft}
            L.~D.~Faddeev, L.~A.~Takhtadzhan,
            {\it The quantum method of the inverse problem and the Heisenberg $XYZ$ model}, Rus. Math. Surv., \textbf{34} (1979), No.~5, 11--68.

\bibitem{bt}
         N.~Bogoliubov, J.~Timonen,
         {\it Correlation functions for a strongly coupled boson system  and
         plane partitions}, Phil. Trans. of Royal Soc. A, \textbf{369} (2011), No.~1939, 1319--1333.

\end{thebibliography}
